\documentclass[preprint,12pt]{elsarticle}

\usepackage{amssymb}
\usepackage{amsmath}

\usepackage{lineno}

\usepackage{booktabs}
\usepackage{hyperref}
\usepackage{algorithm}
\usepackage{algorithmic}
\usepackage{caption}
\usepackage{subcaption}
\usepackage{float}
\usepackage{placeins}
\usepackage{afterpage}
\usepackage{balance}

\journal{Elsevier}

\begin{document}

\begin{frontmatter}



\title{Enhancing deep learning of ammonia/natural gas combustion kinetics via physics-aware data augmentation and scale separation}



\author[1,3]{Ke Xiao}

\affiliation[1]{organization={State Key Laboratory of Turbulence and Complex Systems, School of Mechanics and Engineering Science, Peking University},
            city={Beijing},
            citysep={}, 
            postcode={100871}, 
            country={China}}

\author[1]{Yangchen Xu}

\author[1,3]{Han Li\corref{coa}}
\cortext[coa]{Corresponding author}
\ead{han_li@pku.edu.cn}

\author[1,3]{Zhi X. Chen}


\affiliation[3]{organization={AI for Science Institute (AISI)},
            city={Beijing},
            citysep={}, 
            postcode={100080}, 
            country={China}}


\begin{abstract}
Accurate and efficient numerical simulation of ammonia combustion is critical for advancing ammonia-based energy systems, where turbulent flame dynamics and pollutant formation strongly affect practical applicability. However, such simulations are hindered by the need to solve high-dimensional stiff chemical ordinary differential equations (ODEs), which constitute the primary computational bottleneck. To address this challenge, this study explores {\fontfamily{qcr}\selectfont Deep} learning for solving {\fontfamily{qcr}\selectfont Flame} chemical kinetics with stiff {\fontfamily{qcr}\selectfont ODEs} (DFODE) in ammonia/natural gas combustion. Thermochemical training data are obtained from one-dimensional (1D) freely propagating premixed laminar flames, and a physics-aware augmentation strategy combining interpolation of neighboring states with constrained random perturbations is introduced to overcome sampling imbalance near steep flame-front gradients. In addition, transformation strategies for model target formulation were evaluated, and the prediction accuracy in low-temperature regimes was notably enhanced through scale separation for targets spanning multiple orders of magnitude. Validation in 1D laminar flames confirms the effectiveness of these refinements, while {\em a posteriori} evaluation in a two-dimensional (2D) propagating flame under homogeneous isotropic turbulence (HIT) demonstrates that the trained models generalize to unseen conditions. The DNN surrogates reproduce flame characteristics with high fidelity and deliver up to a $20\times$ speedup in end-to-end CFD simulations. These results highlight the potential of deep learning–based chemical kinetics to accelerate ammonia/natural gas combustion modeling, supporting efficient and scalable high-fidelity simulations for emerging zero-carbon energy systems.
\end{abstract}





\end{frontmatter}


\section{Introduction}\label{introduction}

Ammonia has gained significant attention as a carbon-free fuel and hydrogen carrier \cite{valera-medina_ammonia_2018,valera-medina_review_2021,kojima_ammonia_2022,dimitriou_review_2020,sun_unlocking_2025}, aligning with the United Nations' Sustainable Development Goals (SDGs), which aim to pursue sustainable energy solutions that are accessible, reliable, and environmentally friendly. Key benefits of ammonia as a fuel include its zero carbon content, high volumetric energy density, and an established infrastructure for production, storage, and transport \cite{yapicioglu_review_2019,aziz_ammonia_2020}. However, challenges such as low flammability, low radiation intensity, and high NO$_\text{x}$ emissions due to its nitrogen content necessitate further research to enhance ammonia combustion. Significant efforts have focused on blending ammonia with highly reactive fuels, such as hydrogen, natural gas \cite{juric_emission_2025}, diesel \cite{wu_effects_2025}, and dimethyl ether, to improve overall combustion properties like burning velocity and adiabatic flame temperature, thus enabling the use of ammonia in existing combustion systems with minimal modifications \cite{chai_review_2021,ku_propagation_2020}.

Numerical simulations are crucial for understanding the complex interactions in ammonia-based fuel blends and optimizing their combustion characteristics. To achieve this, detailed finite-rate chemistry models are essential, as they provide more fundamental insights into combustion phenomena, particularly for ammonia-fueled systems, where flame instability dynamics can be pronounced \cite{lu_toward_2009,liu_revealing_2025}. However, incorporating detailed chemical kinetics significantly increases computational costs, especially in simulations involving ammonia blends, where comprehensive chemical mechanisms typically consist of dozens of species \cite{mikulcic_numerical_2021,honzawa_predictions_2019}. Description of chemistry using these mechanisms requires direct integration of stiff systems of ordinary differential equations (ODEs) that govern species concentration evolution. The stiffness, stemming from the wide range of chemical timescales, necessitates extremely small time steps for stable and accurate integration, thereby making the process highly computationally demanding. In practical simulations, chemistry evaluation alone can account for over 90\% of total computational time, posing a major bottleneck in high-fidelity combustion modeling.


Recently, machine learning techniques for accelerating chemistry-related computations have gained increasing interest \cite{meuwly_machine_2021,zhang_multi-scale_2022,echekki_machine_2023,ihme_artificial_2024}. Deep learning, a subset of machine learning utilizing deep neural networks (DNNs) to model complex patterns in large datasets, has been applied in combustion simulations as an alternative to traditional ODE integrators. By framing the integration process as a regression problem—predicting changes in species concentrations over time based on initial conditions—deep learning models can approximate chemical source terms without the need for stiff ODE integration at each time step. This approach can also be viewed as a form of data storage and retrieval or tabulation, as DNNs learn patterns from extensive datasets of precomputed chemistry calculations and perform inference with new, unseen data during simulations \cite{ren_numerical_2014, ding_machine_2021}.

Compared to traditional tabulation methods, DNNs avoid the need for dimension reduction, which can compromise accuracy. They typically require significantly less memory \cite{nikolaou_criteria_2022}, since they only need to store a set of model parameters and weights instead of input-output pairs. Furthermore, DNN inference benefits from modern Graphics Processing Units (GPUs), optimized for parallel computation, which enables rapid processing of multiple inputs simultaneously. This greatly enhances inference speed and makes real-time evaluations feasible in complex chemical systems. However, challenges such as data curation, model selection, evaluation, and validation remain significant hurdles, limiting the broader application of this methodology.


Addressing these hurdles begins with the creation of high‑quality training datasets, since the effectiveness and reliability of DNN‑based chemistry models directly depend on the representativeness of their data. Due to the limited extrapolation capabilities of data-driven methods, researchers have focused on constructing datasets that encompass thermochemical states representative of those encountered in later combustion simulations. For instance, An et al. \cite{an_artificial_2020} sampled data from a Reynolds-Averaged Navier-Stokes (RANS) pre-simulation and applied their models in Large Eddy Simulations (LES) of the same problem, while Chi et al. \cite{chi_--fly_2021} utilized on-the-fly training in Direct Numerical Simulations (DNS) to ensure that the training data is representative of inference data.

While problem‐specific sampling from the target case ensures high fidelity, it often limits the generalizability and reusability of DNN-based chemistry models across diverse scenarios. To address this, researchers have adopted generic canonical problem sampling, leveraging fundamental reacting‐flow configurations to generate broadly representative thermochemical datasets. For example, Wan et al. \cite{wan_chemistry_2020} sampled data from turbulent, non-adiabatic, non-premixed micro-mixing canonical problems and applied the trained neural networks to DNS of a syngas turbulent oxy-flame. Similar configurations were also used by Béroudiaux et al. \cite{beroudiaux_artificial_2025} for modeling high-pressure hydrogen–air combustion. Chatzopoulos and Rigopoulos \cite{chatzopoulos_chemistry_2013} extracted data from non-premixed flamelets at varying strain rates for RANS-PDF simulations of CH$_4$/H$_2$/N$_2$ turbulent flames, a low-dimensional manifold sampling methodology further expanded by Franke et al. \cite{franke_tabulation_2017}, Ding et al. \cite{ding_machine_2021}, and Readshaw et al. \cite{readshaw_simulation_2023}. Such approaches have also been explored to create initial databases for other storage and retrieval methods by Newale et al. \cite{newale_feasibility_2022}. They demonstrated that both flamelets and partially stirred reactor (PaSR) simulations can effectively generate representative thermochemical states encountered in combustion simulations. The core idea behind these generic canonical problem sampling methods is that, under the same initial working conditions, ensembles of thermochemical states throughout the evolution of chemical systems form a continuous manifold \cite{maas_simplifying_1992}.


Despite recent progress, the data preparation process for DNN-based chemical kinetics remains a non-trivial challenge. Constructing large, high-quality datasets that accurately capture the complex thermochemical states encountered in reactive flows is critical yet difficult, particularly for fuels with intricate chemical behavior. Notably, most existing studies in this area have concentrated on hydrocarbon fuels with relatively simplified mechanisms. In contrast, ammonia combustion introduces additional complexity due to nitrogen-bearing species and a wide spectrum of chemical time scales, which significantly complicate both data generation and model learning.

Accurately predicting species concentrations in combustion chemistry modeling also presents a significant challenge due to their inherently multiscale nature, with target values spanning several orders of magnitude. This wide dynamic range complicates the training of deep learning models, as standard approaches often struggle to maintain predictive accuracy across such disparate scales. Addressing this issue requires strategies that effectively manage the multiscale characteristics of the data, ensuring that models can produce dependable predictions across the entire concentration spectrum without being hindered by the extreme disparities in magnitude.

Previous studies have explored various strategies to address the challenges of modeling multiscale targets in combustion chemistry. One notable approach is the Multiple Multilayer Perceptron (MMLP) method proposed by Ding et al. \cite{ding_machine_2021}, which employs ensemble learning by training multiple specialized neural networks for different data scales, thereby improving accuracy within specific magnitude regions. However, this approach faces limitations as the benefits diminish with further data partitioning, and the increasing complexity of managing numerous models becomes a drawback. Another common strategy involves applying nonlinear transformations, such as the Box-Cox \cite{zhang_multi-scale_2022,zhang_graphics_2024,li_comprehensive_2025} or log-transformations \cite{beroudiaux_artificial_2025}, to the data to reduce the disparity in target magnitudes, aiming to improve model stability and accuracy across the multiscale range.


To address these gaps, this study explores deep learning for solving flame chemical kinetics with high-dimensional stiff ordinary differential equations (DFODE) for ammonia/natural gas combustion using the comprehensive mechanism by Okafor et al.~\cite{okafor_experimental_2018}, comprising 59 species and 356 elementary reactions. Thermochemical training data are generated from canonical one-dimensional premixed laminar flames, and a physics-aware data augmentation strategy is introduced to improve coverage in underrepresented regions, particularly near steep thermochemical gradients. In addition, transformation strategies for model target formulation and their effect on prediction accuracy across targets of several orders of magnitude were evaluated. The trained models are first validated in 1D laminar flames and subsequently evaluated {\em a posteriori} in a two-dimensional homogeneous isotropic turbulence (HIT) flame configuration to assess generalizability and computational performance. This study advances the application of deep neural networks to ammonia-fueled combustion and presents a scalable framework for accelerating high-fidelity simulations involving complex chemical kinetics.

\section{Methodology}
\subsection{Preliminaries on Deep Learning for Combustion Chemistry Integration}
A chemical system with $N$ species and $M$ reactions can be expressed as:
\begin{equation}
    \sum_{\alpha=1}^{N} \nu'_{\alpha j} \mathcal{M}_{\alpha} \Rightarrow \sum_{\alpha=1}^{N} \nu''_{\alpha j} \mathcal{M}_{\alpha} \quad \text{for } j = 1, \ldots, M, \label{eq:chem_system}
\end{equation}
Here, \( \mathcal{M}_{\alpha} \) denotes species \( \alpha \), while \( \nu'_{\alpha j} \) and \( \nu''_{\alpha j} \) represent the molar stoichiometric coefficients for species \( \alpha \) in reaction \( j \). The reaction rate of species \( \alpha \) is described by the following ODE:
\begin{equation}
    \frac{\text{d} Y_{\alpha}}{\text{d}t} = W_{\alpha} \sum_{j=1}^{M} \left( \nu'_{\alpha j} - \nu''_{\alpha j} \right) \\
    \left\{ K_{fj} \prod_{\alpha=1}^{N} \left( \frac{\rho Y_{\alpha}}{W_{\alpha}} \right)^{\nu'_{\alpha j}} - 
    K_{rj} \prod_{\alpha=1}^{N} \left( \frac{\rho Y_{\alpha}}{W_{\alpha}} \right)^{\nu''_{\alpha j}} \right\}
    \label{eq:chem_ode}
\end{equation}
where \( Y_{\alpha} \) is the mass fraction and \( W_{\alpha} \) is the molecular weight of species \( \alpha \). The mixture mass density is represented by \( \rho \), while \( K_{fj} \) and \( K_{rj} \) are the forward and reverse rate constants for reaction \( j \). These rate constants are typically parameterized by temperature \( T \) according to Arrhenius models and also pressure \( p \) for pressure-dependent reaction rates.

For simulation methods that require real-time computation of the chemical source term in every grid cell and time step, the source terms \( \dot{\omega}_\alpha \) are obtained through time integration of the ODE system mentioned above. This process could be replaced by deep learning surrogate models that fit the nonlinear mapping functions between input-output pairs.

\begin{figure}[h]
	\centering 
	\includegraphics[width=0.98\textwidth]{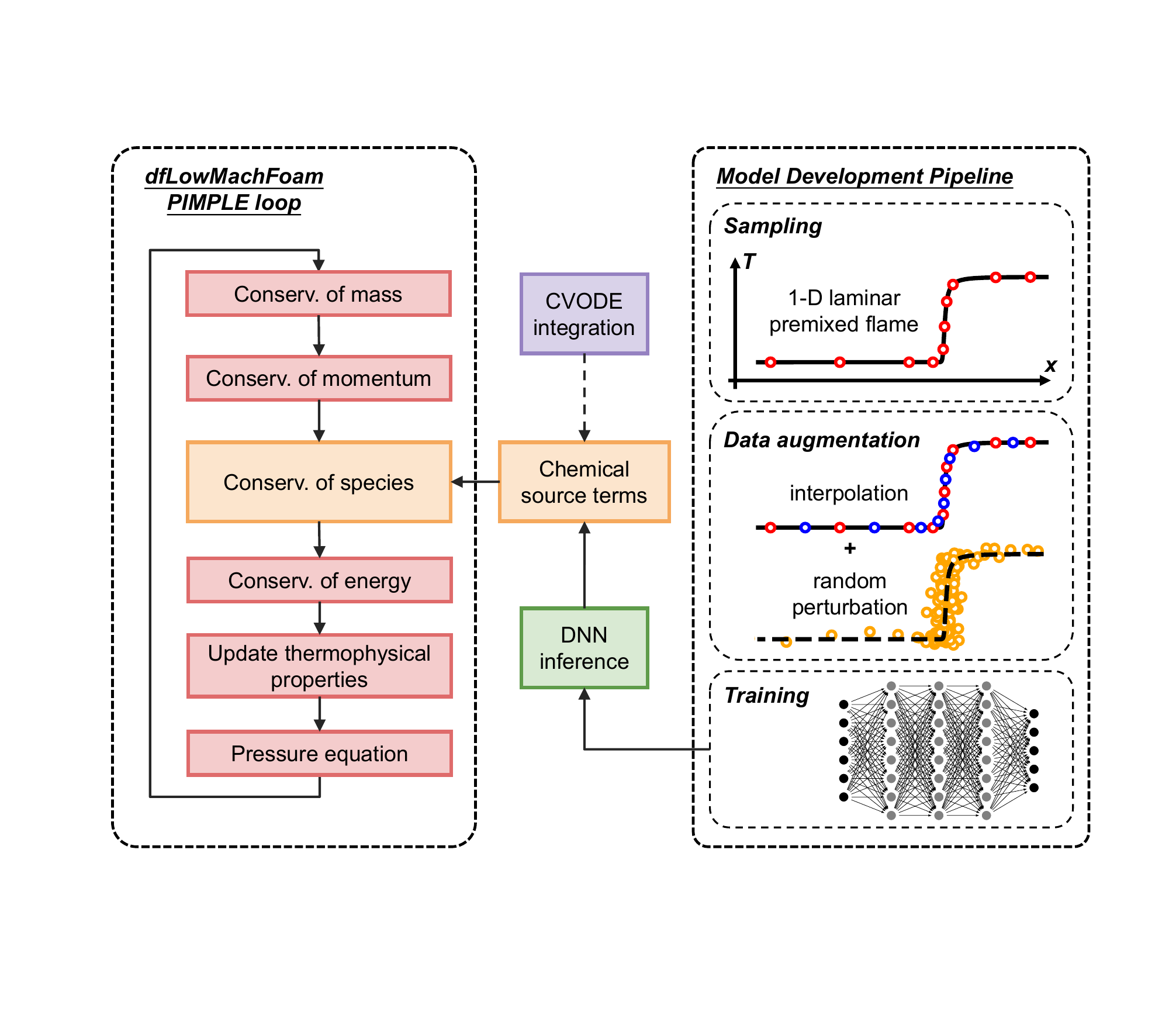}	
	\caption{Overview of the training procedure and flow chart for the dfLowMachFoam solver, illustrating the integration of DFODE models to provide chemical source terms.}
	\label{fig:flow_chart}
\end{figure}

As shown in Equation \ref{eq:chem_ode}, in combustion processes, the instantaneous reaction source terms \( \dot{\omega}_\alpha \) can be expressed as a function of \( T \), \( p \) and the mass fraction vector \( \mathbf{Y} = (Y_1, Y_2, ... , Y_N) \). Therefore, the inputs chosen for the neural networks are local thermochemical parameters \( \{ T(t), p(t), \mathbf{Y}(t) \} \), and the outputs are the species increments \( \{ \mathbf{Y}(t + \Delta t) - \mathbf{Y}(t) \} \). In this work, a time step \( \Delta t = 1 \times 10^{-6} \)s is adopted to define these increments. The predicted outputs are subsequently used to calculate chemical source terms. Figure \ref{fig:flow_chart} illustrates the training process of DFODE models and their integration with reacting flow solvers. The training process starts with the collection of input datasets, which are then fed into the DFODE architecture consisting of an input layer, multiple hidden layers, and an output layer. Throughout the training phase, the DFODE develops the ability to accurately map input parameters to desired outputs. This is achieved through the optimization of a loss function that measures the difference between the predicted outputs and the actual results generated by the CVODE integrators from SUNDIALS, as provided by Cantera \cite{goodwin_cantera_2022}.

Once trained, the DFODE is coupled with the dfLowMachFoam solver from the open-source reacting flow simulation platform DeepFlame \cite{mao_deepflame_2023,mao_integrated_2024,mao_deepflame_2025}. This solver computes the flow field and local conditions of the reacting mixture. At each time step, dfLowMachFoam retrieves the current simulation values and inputs them into the DFODE for inference. The DFODE then predicts reaction rates, which are incorporated into species conservation equations.

\subsection{Data Collection from Canonical Flame}
Data-driven models often face challenges in generalizing beyond their training domain, especially within the complex, high-dimensional space of combustion chemistry. Ensuring that the training dataset comprehensively covers the relevant thermochemical states encountered during turbulent combustion is essential for developing robust models suitable for CFD simulations.

Previous research in deep learning for combustion chemistry integration has predominantly relied on datasets derived from small-scale, computationally efficient simulations of canonical combustion problems \cite{wan_chemistry_2020,ding_machine_2021}. These datasets typically encompass thermochemical states that resemble those found in realistic turbulent flames with similar fuel and oxidizer compositions. Such an approach provides a practical foundation for training surrogate models, as it captures key features of the relevant high-dimensional thermochemical composition space while maintaining computational tractability.

Many prior studies have aimed to develop neural network models with broad generalization capabilities by combining data from multiple small-scale simulations with different configurations \cite{ding_machine_2022,li_comprehensive_2025}. The rationale is to ensure diverse thermochemical state coverage across various operating conditions, such as different turbulence intensities, equivalence ratios, or fuel blending ratios. This approach helps to enhance the robustness and applicability of the surrogate models across a wide range of turbulent combustion scenarios.

In contrast, this work focuses on identifying the minimal data requirements for effective model performance in CFD simulations by limiting data generation to a single canonical combustion problem. This approach streamlines data collection, accelerates data generation, and facilitates easier updates or expansions of the dataset. It also simplifies the analysis of dataset quality and helps to understand which factors influence model accuracy and generalization. While the applicability of the trained model is limited to turbulent flames under the same operating conditions as the sampled canonical flame, the data generation process can be readily adapted to different blending ratios or mixture compositions, thereby extending the methodology’s overall applicability.

In this work, we develop DFODE models tailored for premixed combustion of ammonia/natural gas blends with a specific NH$_3$ blending ratio of 60\% and a stoichiometric equivalence ratio. The choice of a higher ammonia blending ratio is driven by ongoing research interest in ammonia-rich combustion conditions, which remain relatively underexplored in the context of deep learning-based combustion chemistry modeling. Our goal is to enable these models to accurately simulate NH$_3$/CH$_4$ premixed flames under turbulent flow conditions.

To achieve this, we sample from a single, 1D freely propagating premixed laminar flame of premixed 60\%NH$_3$/40\%CH$_4$ mixture with an equivalence ratio of 1. The unburnt gas temperature is set to 300 K, and the ambient pressure is 1 atm. The computational domain includes a premixed fuel/air mixture on one side and equilibrium states computed via Cantera on the other. The mesh consists of 500 cells, resolving the flame thickness with 10 grid points, and the inlet velocity is determined by the laminar flame speed to sustain the flame front. Simulations are conducted with a time step of $\Delta t = 1 \times 10^{-6}$ seconds over $2.5 \times 10^{-3}$ seconds, sampling data at every time step. This process generates an initial dataset of approximately 1,250,000 states.

\subsection{Flame Structure Interpolation to Mitigate Data Imbalance}
Effective modeling of combustion chemical kinetics through deep learning regression techniques critically depends on the availability of representative and diverse datasets. However, data imbalance, in which certain regions of the composition space are overrepresented while others are sparsely sampled, presents a fundamental challenge that can significantly impair model accuracy and generalization \cite{yang_delving_2021}. This issue is particularly pronounced in combustion chemistry, where datasets are often dominated by slowly reacting states near equilibrium or in cold mixtures, characterized by minimal thermochemical variations. Conversely, reactive regions such as the flame front, which exhibit rapid changes in temperature and species concentrations, tend to be underrepresented. This imbalance can limit a model's ability to accurately predict chemical kinetics for reacting regions, which is crucial for reliable simulations that capture dynamic flame behavior and represent a wide spectrum of combustion phenomena.

Previous research has acknowledged the challenge of data imbalance in combustion datasets, and several strategies have been proposed to address it. For instance, Saito et al. \cite{saito_data-driven_2023} addressed this issue by employing a weighted sampling approach based on kernel density estimates of the data distribution, thereby ensuring a more uniform representation across reaction rate regions during model training. Chen et al. \cite{chen_clustering-enhanced_2023} utilized CVODE’s adaptive time stepping to generate a more balanced dataset, with denser sampling in fast reaction regions within zero-dimensional reactor simulations. Wan et al. \cite{wan_chemistry_2020} artificially enriched the training dataset with copies of original thermochemical states in which the H$_2$O$_2$ mass fraction is replaced by a random value between zero and its peak level. This helped to enable the neural network to accurately learn the species' rapid evolution. Additionally, data clustering techniques have also been extensively explored within the context of deep learning for combustion chemistry integration \cite{blasco_self-organizing-map_2000,han_improved_2022,chen_clustering-enhanced_2023} and may help mitigate data imbalance. While these methods contribute valuable strategies for alleviating data imbalance, a systematic and physically consistent augmentation of the dataset remains an open challenge. Specifically, generating physically meaningful states within reactive zones is essential for improving model robustness and predictive capability.

\begin{figure}[t]
    \centering
    \includegraphics[width=0.98\textwidth]{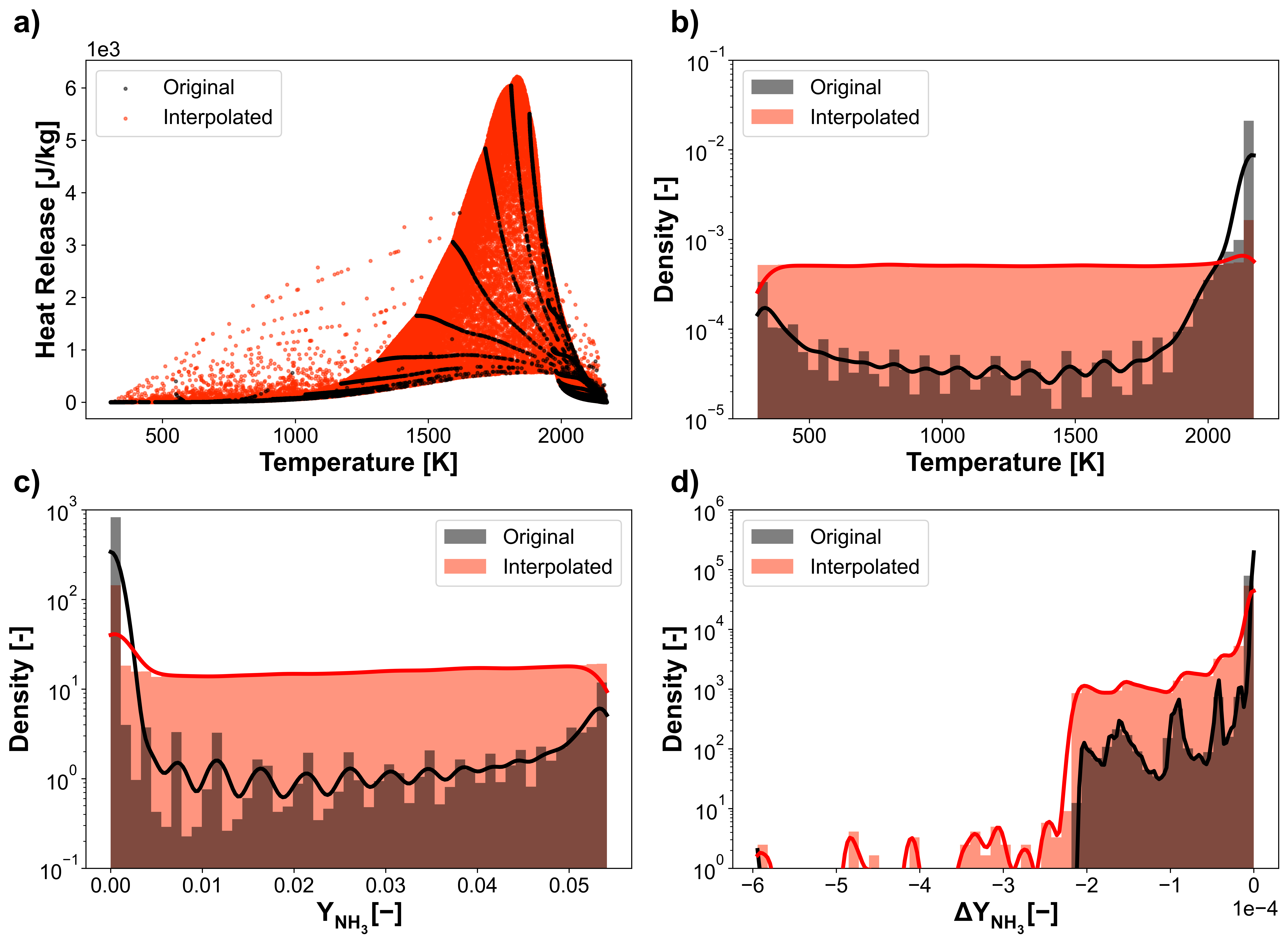}
    \caption{
    Comparison of thermochemical data before and after flame structure interpolation. a) Scatter plot of heat release over reaction step versus initial temperature. b-d) Density distribution of temperature, ammonia mass fraction $Y_{\text{NH}_3}$, and ammonia concentration change $\Delta Y_{\text{NH}_3}$ for original and interpolated datasets.
    }
    \label{fig:interpolation_histogram}
\end{figure}

Building upon this concept, the present work introduces a data augmentation procedure based on flame structure interpolation to address data imbalance. Focusing on one-dimensional laminar premixed flames, where the flame front exhibits pronounced temperature gradients, the proposed method aims to generate additional thermochemical states that are physically consistent and representative of the reacting zone. The detailed interpolation procedure is outlined in Algorithm \ref{alg:flame-structure-interp}, which involves collecting thermochemical states within each simulation timestep, sorted according to their spatial position. Linear interpolation is then performed between existing states at evenly spaced temperature intervals, thus producing a more evenly distributed dataset across the temperature spectrum.

Figure \ref{fig:interpolation_histogram} illustrates the impact of this interpolation-based augmentation on data distribution. The histograms compare the density distribution of temperature, ammonia mass fraction $Y_{\text{NH}3}$, and the changes in ammonia concentration $\Delta Y_{\text{NH}_3}$ over \(\Delta t\) for datasets with and without interpolation. The augmented dataset demonstrates a markedly more uniform distribution across all parameters, effectively balancing sampling frequencies and alleviating data imbalance. This enhanced representation enables deep neural network models to better generalize across input ranges, particularly in regions where data were previously sparse, thereby improving the reliability of predictions in highly reactive combustion zones.

\begin{algorithm}
\caption{Data Augmentation through Linear Interpolation}
\label{alg:flame-structure-interp}
\begin{algorithmic}
\STATE \textbf{Input:} Thermochemical dataset from 1D laminar flame simulation, sorted by spatial position in 1D domain (includes temperature \( T \), pressure \( p \), and species mass fractions \( \mathbf{Y} \))
\STATE \textbf{Input:} A grid of evenly spaced temperature values within \([T_{\text{min}}, T_{\text{max}}]\) of original dataset: \( T_{\text{grid}} = [T_{\text{min}}, T_{\text{min}} + \Delta T, T_{\text{min}} + 2\Delta T, \ldots, T_{\text{max}}] \)

\FOR{each adjacent pair \( (T_i, p_i, \mathbf{Y}_i) \) and \( (T_{i+1}, p_{i+1}, \mathbf{Y}_{i+1}) \) in the dataset}
    \FOR{each \( T_{\text{new}} \) in \( T_{\text{grid}} \)}
        \IF{\( T_i < T_{\text{new}} < T_{i+1} \)}
            \STATE Compute the interpolation for pressure \( p_{\text{new}} \):
            \[
            p_{\text{new}} = p_i + \frac{(p_{i+1} - p_i)}{(T_{i+1} - T_i)} \cdot (T_{\text{new}} - T_i)
            \]
            \STATE Compute the interpolation for species mass fractions \( \mathbf{Y}_{\text{new}} \):
            \[
            \mathbf{Y}_{\text{new}} = \mathbf{Y}_i + \frac{(\mathbf{Y}_{i+1} - \mathbf{Y}_i)}{(T_{i+1} - T_i)} \cdot (T_{\text{new}} - T_i)
            \]
            \STATE Store \( (T_{\text{new}}, p_{\text{new}}, \mathbf{Y}_{\text{new}}) \)
        \ENDIF
    \ENDFOR
\ENDFOR

\STATE \textbf{Output:} Augmented dataset containing both original and interpolated thermochemical states
\end{algorithmic}
\end{algorithm}

Another issue is that the thermochemical states derived from canonical flames may not fully capture the relevant composition space required for turbulent combustion simulations. They also follow well-defined trajectories within the sample space, making the trained model susceptible to perturbations or deviations from this manifold. To enhance model robustness, data augmentation through random perturbations can simulate multi-dimensional transport and turbulence effects on the thermochemical structure of the flame, thereby achieving broader coverage of the composition space \cite{ding_machine_2021}. In this approach, the original data are discarded in favor of the randomly perturbed data for training, enabling the DFODE to learn from points beyond the lower-dimensional flames and improving its applicability to new states.

\begin{figure}[p]
    \centering
    \includegraphics[width=0.98\textwidth]{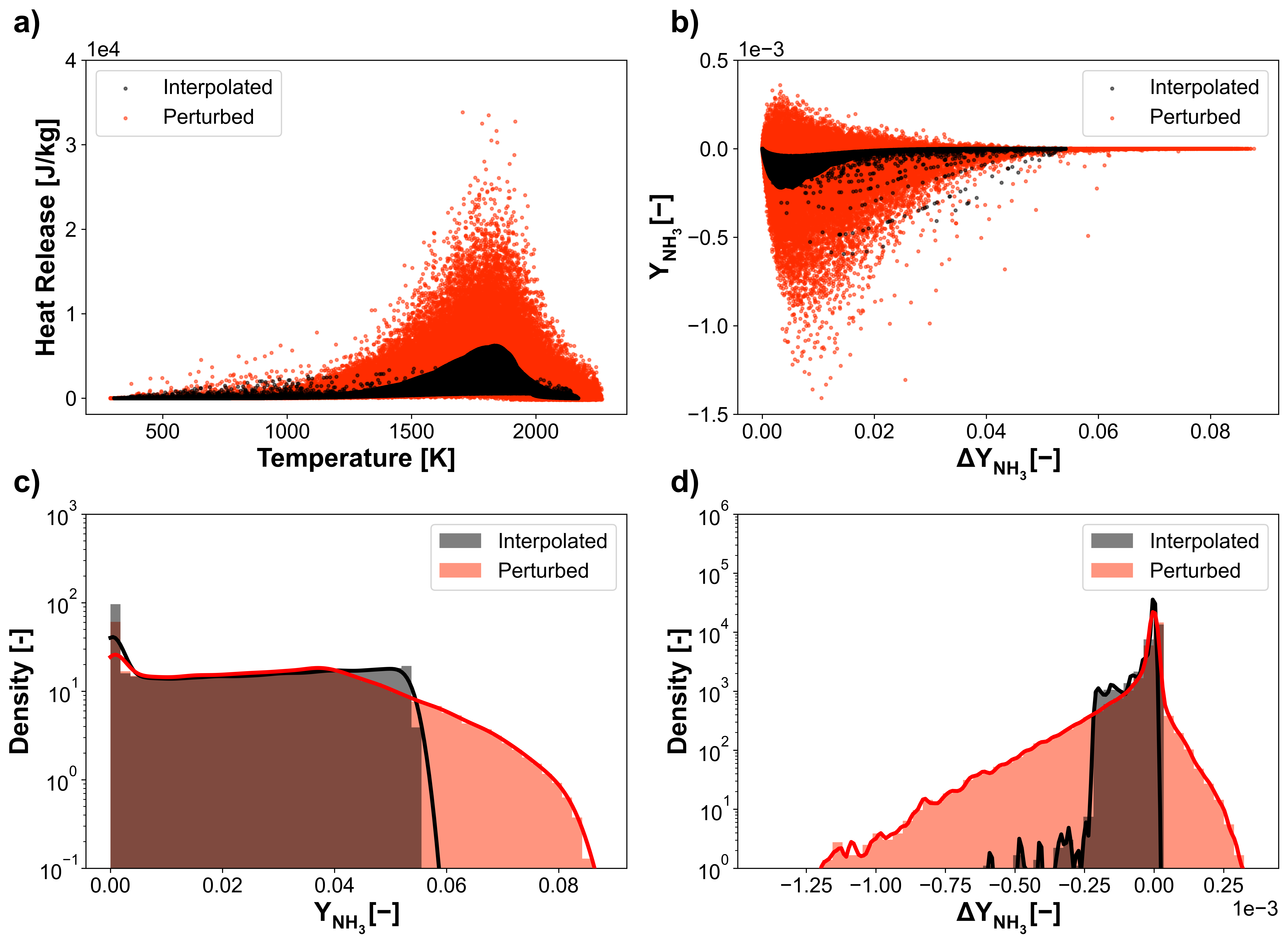}
    \caption{
    Comparison of interpolated thermochemical data before and after perturbation augmentation. a) Scatter plot of heat release over reaction step versus initial temperature. b) Scatter plot of initial ammonia mass fraction versus ammonia concentration change. c,d) Density distribution of ammonia mass fraction $Y_{\text{NH}_3}$ and ammonia concentration change $\Delta Y_{\text{NH}_3}$ for interpolated and perturbed datasets.
    }
    \label{fig:perturbation_scatter}
\end{figure}

In this work, both the initial and interpolation-generated states undergo random perturbation defined by:
\begin{align}
    T' &= T + 100 \cdot X \label{pert_t} \\
    p' &= p + (\text{max}(p) - \text{min}(p)) \cdot 0.15 \cdot X \label{pert_p} \\
    Y_\alpha' &= Y_\alpha^{1 + 0.15 \cdot X} \label{pert_y}
\end{align}
where \(T', p', Y_\alpha'\) represent the perturbed values, and \(X\) is a random number uniformly distributed between -1 and 1. Figure \ref{fig:perturbation_scatter} shows the composition space spanned by datasets with and without perturbation augmentation. The coverage of the composition space is broadened through random perturbation. However, if left unconstrained, it may produce numerous nonphysical states, adversely affecting model accuracy in critical regions. Such nonphysical states are primarily observed in areas with higher initial temperatures, corresponding to near-equilibrium states, and demonstrate significant negative heat release rates.

To address these issues, nonphysical states are filtered based on specific criteria. Perturbed temperatures must satisfy the condition that they lie within the range of:
\[ 
290 \leq T' \leq \text{max}(T) + 100 
\]
where \(\text{max}(T)\) is the maximum temperature from the unperturbed dataset. Additionally, the mass fraction of \( Y_{N_2}' \) must fall within the limits defined by:
\[
Y_{\text{min}} - 0.05 \cdot \text{span}(Y_{N_2}) \leq Y_{N_2}' \leq Y_{\text{max}} + 0.05 \cdot \text{span}(Y_{N_2})
\]
where \( \text{span}(Y_{N_2}) = Y_{\text{max}} - Y_{\text{min}} \). Perturbed states must also adhere to limits on heat release rates, with states exhibiting significant negative heat release rates discarded to ensure physical realism. The perturbation process is performed over several iterations, during which perturbed data is collected. After applying the filtering criteria, approximately 8,000,000 valid perturbed states are randomly selected to form the final training dataset.

\subsection{Scale-separation Transformation for Multiscale Targets}
In combustion chemistry modeling, the changes of species concentrations during the reaction step often span several orders of magnitude, as illustrated in Figures 1 and 2, where species concentrations can range from extremely small values such as \(10^{-25}\) to relatively larger ones like \(10^{-5}\). This multiscale nature presents a significant challenge for deep learning models by making it difficult to achieve satisfactorily accurate predictions across targets with disparate scales.

\begin{figure}[h]
    \centering
    \includegraphics[width=0.98\textwidth]{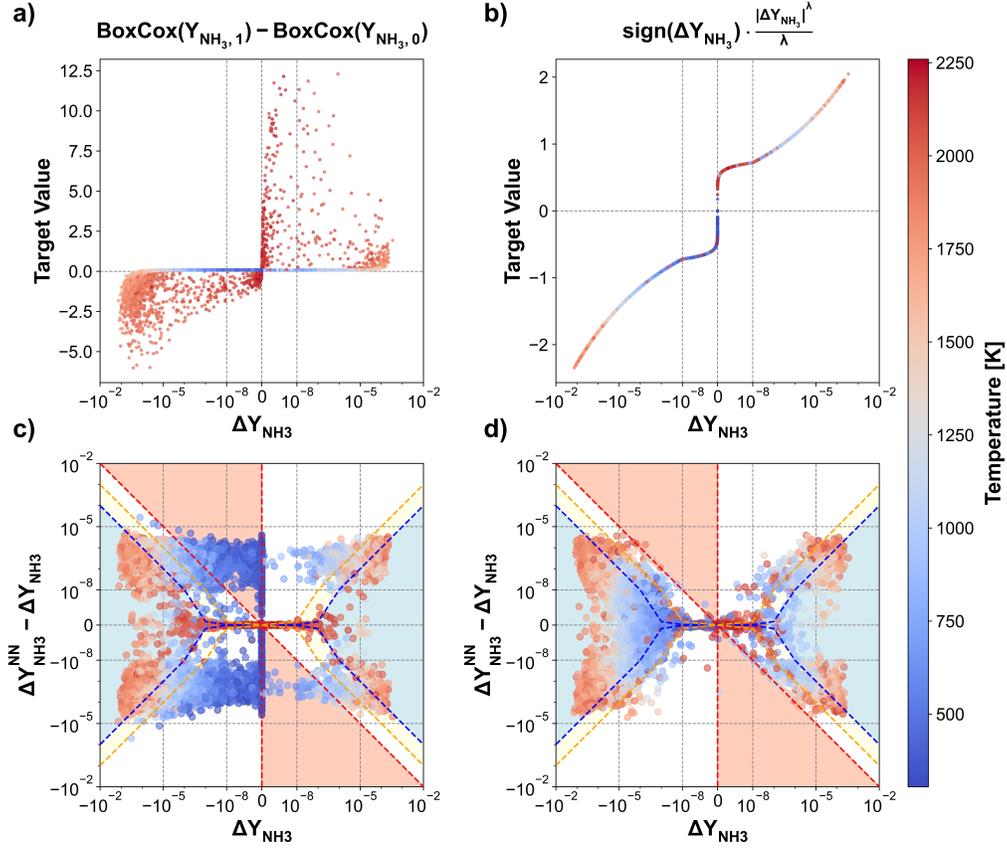}
    \caption{
    Comparison of target value formulations and model performance colored by temperature. a) \& c): Box-Cox transformation. b) \& d): Power transformation. Top row: Ground truth ammonia concentration change versus formulated target values used for training. Bottom row: Ground truth versus model prediction error. For all x-axes and y-axes of c) and d) panel, values with magnitude below \( 10^{-8} \) are represented on a linear scale, while values exceeding \( 10^{-8} \) are depicted on a logarithmic scale.
    }
    \label{fig:pairity_plot_transform}
\end{figure}

Ensuring accuracy across targets with disparate scales requires the model to produce dependable predictions for a wide range of concentration magnitudes, from extremely small to relatively large values. For targets with larger magnitudes, such as in the range of \(10^{-6}\) to \(10^{-3}\), the goal is to achieve smaller relative errors to ensure high predictive fidelity. However, as the target magnitude decreases, the usefulness of relative error as a metric diminishes. For instance, a prediction of \(10^{-20}\) when the true concentration is \(10^{-25}\) results in a relative error of \(10^{5}\), which appears large but may not be practically problematic given the extremely small scale involved. In this context, a prediction is considered acceptable if it remains within a very small magnitude, regardless of the relative error. Consequently, in this study, dependable predictions for relatively larger targets are characterized by low relative error, whereas for extremely small concentrations, accuracy is regarded as sufficient if predictions are also of very small magnitude, preserving their physical relevance.

Previous research has investigated various strategies to address the challenges posed by multiscale targets. A prominent example is the Multiple Multilayer Perceptron (MMLP) method proposed by Ding et al. \cite{ding_machine_2021}. The MMLP approach is grounded in ensemble learning principles, wherein multiple simple neural networks are trained to specialize in data corresponding to different scales. The selection of the appropriate network for inference can be based on either the output range or the input range. In essence, this method involves partitioning the data according to the magnitude of inputs or outputs and training separate networks tailored to each partition. The primary advantage of this strategy is that it localizes prediction accuracy within specific magnitude regions, thereby avoiding the degradation of performance that can occur when a single network attempts to model data spanning a wide dynamic range. However, as acknowledged by the authors in \cite{ding_improved_2025}, this partitioning process cannot be extended indefinitely. Specifically, as the data output range narrows, the marginal gains in prediction accuracy from training additional networks diminish, and may even become detrimental. Furthermore, the reduction in available data within increasingly small partitions renders further subdivision redundant. An additional limitation of this approach is the significant increase in complexity associated with model training, evaluation, and management. For instance, in their most recent work \cite{ding_improved_2025} involving the GRI-3.0 mechanism, which includes 52 species and exhibits a level of complexity comparable to that of the current study, a total of 161 MLPs were trained, with each species represented by between one and five networks.

Another approach involves applying nonlinear transformations, such as the Box-Cox transformation \cite{zhang_multi-scale_2022,li_comprehensive_2025} or log-transformation \cite{beroudiaux_artificial_2025}, to the data to normalize their magnitudes. When learning transformed values, a common way formulation for model target is \(\mathcal{F} (\mathbf{Y}(t + \Delta t)) - \mathcal{F} (\mathbf{Y}(t))\) \cite{li_comprehensive_2025}, where \(\mathcal{F}\) is the nonlinear transformation. Figure \ref{fig:pairity_plot_transform} shows scatters of absolute errors of \(\Delta \mathbf{Y}\) predictions against the magnitude of true values for models trained on model target defined as
\begin{multline}
    \hat{\mathbf{Y}} = \text{Box-Cox}(\mathbf{Y}(t + \Delta t)) - \text{Box-Cox}(\mathbf{Y}(t)), \\
    \text{where} \quad \text{Box-Cox}(y) = \frac{y^\lambda - 1}{\lambda}, \quad \lambda=0.1.
    \label{eq:BCT}
\end{multline}
It is evident that for targets with magnitudes smaller than 10$^{-8}$, it is not guaranteed that the predictions are also of this magnitude range, signaling undependable predictions in this region. To address this, we propose reformulating the target as the transformed concentration change:
\begin{multline}
    \hat{\mathbf{Y}} = \mathcal{F}[\mathbf{Y}(t + \Delta t) -\mathbf{Y}(t)], \\
    \text{where} \quad \mathcal{F}(y) = \operatorname{sign}(y) \cdot  \frac{\left| y \right|^\lambda}{\lambda}, \quad \lambda=0.1.
    \label{eq:new_transformation}
\end{multline}
This approach applies the nonlinear transformation directly to the concentration difference, reducing the disparity in target magnitudes—especially for very small values. Figure \ref{fig:pairity_plot_transform} displays scatter plots of absolute prediction errors against true value magnitudes for models trained with this reformulated target. The results show that for targets with magnitudes below 10$^{-8}$, it becomes much less likely for predictions to exhibit large magnitudes. This reformulation effectively decouples the scales of the target variables by emphasizing concentration changes rather than absolute concentrations, thereby improving the robustness and reliability of predictions in multiscale regimes. Further details on model accuracy assessment and strategies for enhancing simulation performance with these reformulated targets will be provided in the results section.

The DFODE model employed in this study is structured as a multilayer perceptron (MLP). It consists of an input layer, an output layer, and four hidden layers, each containing 800 neurons. The mass fraction of the inert species argon is excluded from the model outputs, as it remains constant throughout the reaction process and does not necessitate modeling. Both input and output data are normalized using Z-score normalization to facilitate training stability and convergence. The training procedure aligns with methodologies outlined in prior studies \cite{li_comprehensive_2025}.

\subsection{A Posteriori Validation in Turbulent Flame Simulations}
To validate the DFODE models in combustion simulations, simulations of a flame kernel ignition of premixed NH$_3$/CH$_4$ mixture in two-dimensional homogeneous isotropic turbulence (HIT) were conducted. Previous investigations in deep learning for combustion chemistry integration have employed similar simulation setups \cite{li_comprehensive_2025,cai_efficient_2025}.

In the simulation, a square computational domain of \( L \times L = 28 \times 28 \, \text{mm}^2 \) is used, initialized with a premixed stoichiometric mixture composed of 60\% NH$_3$ and 40\% CH$_4$ at 300 K and 1 atm. To ignite the mixture, a circular hot spot with a radius of \( L / 10 \) filled with equilibrium state gases is placed in the center of the domain. The HIT generation approach in \cite{vuorinen_dnslab_2016} is adopted and the fully evolved velocity field is then mapped to the computational domain as the initial flow field. Boundary conditions are set to zero gradient for temperature and species mass fractions, and a non-reflective wave transmissive condition is used for pressure and velocity. The domain is uniformly discretized with 512 × 512 grids to ensure sufficient resolution for both flame and turbulence.

\section{Results}
\subsection{Validation of Data Augmentation and Interpolation Techniques}
In the methodology section, it was demonstrated that the proposed data augmentation strategies, namely flame structure interpolation and random perturbation, improve the coverage and diversity of the training dataset. These enhancements are crucial for training robust neural network models.

In this section, we validate the effectiveness of these techniques through a comparative analysis. Specifically, two DFODE models were trained using datasets that underwent identical perturbation augmentation. The key difference was that one dataset included the additional interpolation-based data augmentation, while the other did not. Both models were trained formulating model output as \(\mathcal{F} (\mathbf{Y}(t + \Delta t)) - \mathcal{F} (\mathbf{Y}(t))\) with \(\mathcal{F}\) being the Box-Cox transformation. Both models were subsequently applied to simulate the same one-dimensional flame configuration used for training data generation. The primary objective was to evaluate whether the inclusion of the interpolation procedure could improve predictive accuracy, particularly in highly reactive zones such as the flame front.

It is important to note that this validation focuses solely on the impact of the interpolation-based data augmentation. Previous studies have established that perturbation augmentation alone is a common and effective technique to improve model robustness \cite{ding_machine_2021,li_comprehensive_2025}. In our current work, models trained on datasets without perturbation augmentation failed to produce reliable results in one-dimensional laminar flame simulations, regardless of whether interpolation was applied. Therefore, the ablation analysis here isolates the effect of the interpolation procedure.

\begin{figure}[h]
    \centering
    \includegraphics[width=0.98\textwidth]{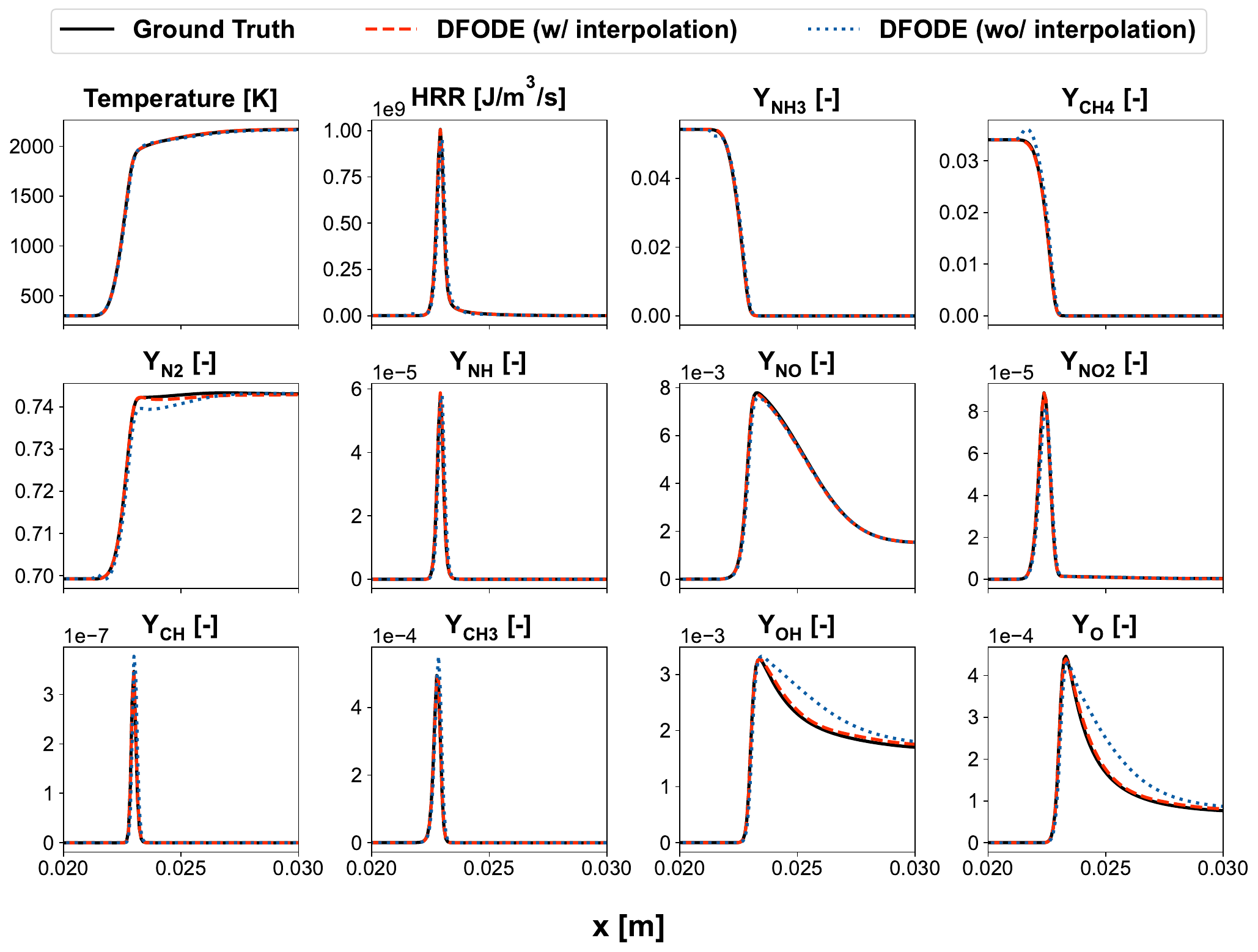}
    \caption{
    Temperature, heat release rate (HRR), and species mass fractions profiles in physical space obtained at 2.5 ms in 1D laminar premixed flame simulations using DFODE inference or CVODE integration for chemical source terms.
    }
    \label{fig:1d_flame_structure_physical_space}
\end{figure}

\begin{figure}[h]
    \centering
    \includegraphics[width=0.98\textwidth]{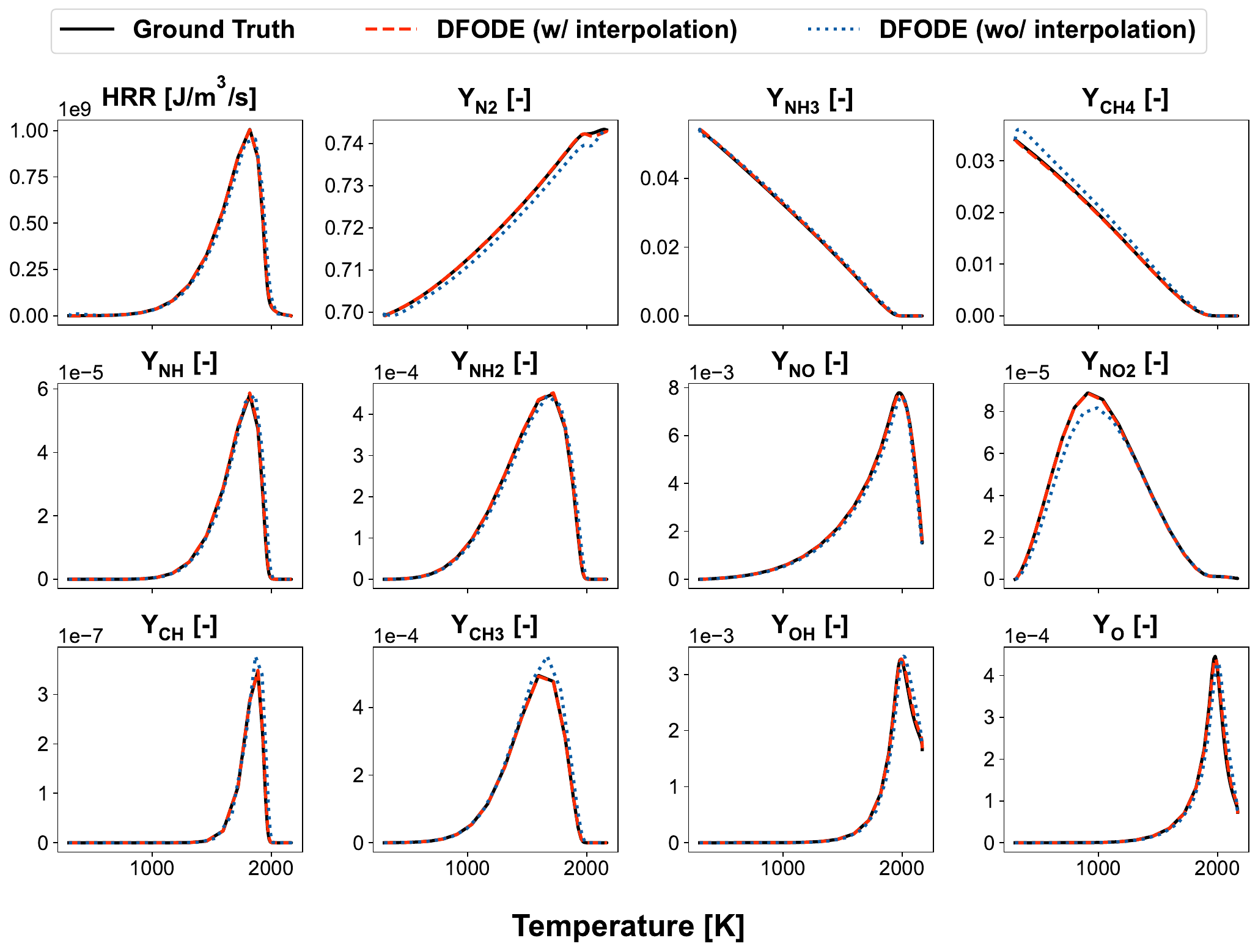}
    \caption{
    Heat release rate (HRR), and species mass fractions profiles in progress variable (temperature) space obtained at 2.5 ms in 1D laminar premixed flame simulations using DFODE inference or CVODE integration for chemical source terms.
    }
    \label{fig:1d_flame_structure_flamelet_space}
\end{figure}

The results, shown in Figure \ref{fig:1d_flame_structure_physical_space}, compare the flame structure at 2.5 ms obtained from the CVODE reference solution and the two DFODE models. Both models captured the temperature profile reasonably well; however, the model trained without interpolation underpredicted the maximum heat release rate. Regarding main species such as NH$_3$, CH$_4$ and N$_2$, the model trained without interpolation exhibited notable discrepancies from the CVODE reference solution. It showed an nonphysical increase in CH$_4$ mass fraction before the flame front and a larger underprediction of N$_2$ behind the flame front. For some key intermediate species, the model without interpolation failed to accurately reproduce the profiles observed in CVODE simulations. Notably, it overpredicted the maximum values for CH and CH$_3$, while underpredicting the consumption of OH and O through reaction behind the flame front. Using temperature as a progress variable, Figure \ref{fig:1d_flame_structure_flamelet_space} shows that these deviations were particularly evident in progress variable space.

In contrast, the DFODE model trained on the dataset that included the interpolation-based augmentation demonstrated significantly improved alignment with the CVODE results. Its predictions of temperature and species mass fractions closely matched the reference solution, capturing the detailed flame structure more accurately. This enhanced performance suggests that the interpolation-augmented dataset provides a more comprehensive representation of the thermochemical space, especially by increasing data density in reactive regions. As a result, the model’s predictive capability is markedly improved.

This validation confirms that incorporating flame structure interpolation into the data augmentation pipeline effectively enhances the neural network’s ability to predict critical thermochemical features within reacting flows. Consequently, this leads to more reliable and physically consistent flame simulations.

\subsection{Evaluation of the Transformation Strategies for Multiscale Targets}
To evaluate the effectiveness of the proposed transformation strategies, two neural network models were trained using identical architecture and optimization hyperparameters, differing solely in the formulation of the target variable \(\Delta \mathbf{Y}\). One model employed the \(
\text{Box-Cox}(\mathbf{Y}(t + \Delta t)) - \text{Box-Cox}(\mathbf{Y}(t))
\) transformation, as used in previous studies \cite{zhang_graphics_2024,li_comprehensive_2025}, hereafter referred to as the BC model. The other utilized a power transformation applied directly to the concentration change, defined as \(
\mathcal{F}[\mathbf{Y}(t + \Delta t) -\mathbf{Y}(t)]
\), where 
\( \quad \mathcal{F}(y) = \operatorname{sign}(y) \cdot  \frac{\left| y \right|^{0.1}}{0.1} \), termed the PT model. Both models were trained on the same dataset, which comprised samples from the 1D laminar flame simulation that underwent both flame structure interpolation and random perturbation augmentations to improve coverage in reactive regions. 

The models' performances were evaluated on two datasets: the training set, consisting of augmented samples from the 1D flame simulation, and a test set derived from a snapshot at 0.004 seconds of a CVODE simulation of the 2D HIT flame. This approach enabled a comparison of each model’s ability to generalize beyond the training data, particularly in capturing the multiscale nature of species concentrations.

To quantitatively assess the model's accuracy in predicting extremely small target values, we introduce the Small-Scale Prediction Index (SSPI). It is defined as the ratio of the number of predictions for which both the predicted value $\left|  Y_{\alpha,i} \right|$ and the true value $\left|  Y_{\alpha,i} \right|$ are below $10^{-15}$, to the total number of true target values below this threshold:
\begin{equation}
\text{SSPI} = \frac{\left| \left\{ i \mid |\hat{Y}_{\alpha,i}| < 10^{-15} \text{ and }\left|  Y_{\alpha,i} \right| < 10^{-15} \right\} \right|}{\left| \left\{ i \mid \left|  Y_{\alpha,i} \right| < 10^{-15} \right\} \right|}
\end{equation}
This metric measures the proportion of predictions that accurately capture the extremely small concentration targets, with an SSPI value of 1 indicating perfect prediction for all targets below the threshold. It offers a direct evaluation of the models’ performance in estimating low-magnitude species concentrations, complementing traditional metrics such as R$^2$ or MAE, which may not fully reflect model performance in this regime.

\begin{table}[htbp]
\centering
\caption{Model Performance Metrics for BC and PT Models}
\label{tab:model_performance}
\begin{tabular}{lcccc}
\toprule
& \textbf{Train} & \textbf{Test} & \textbf{Train} & \textbf{Test} \\
\textbf{Metric} & \textbf{BC Model} & \textbf{BC Model} & \textbf{PT Model} & \textbf{PT Model} \\
\midrule
Num. states & 8,000,000 & 48,916 & 8,000,000 & 48,916 \\
Num. pred. & 464,000,000 & 2,837,128 & 464,000,000 & 2,837,128 \\
MAE (overall) & 5.81e-8 & 5.70e-8 & 1.38e-7 & 1.15e-7 \\
RMSE (overall) & 3.50e-7 & 4.94e-7 & 9.52e-7 & 8.08e-7 \\
$R^2$ (overall) & 0.999977 & 0.999530 & 0.999827 & 0.998741 \\
\midrule
Num. small pred. & 53,183,865 & 467,570 & 53,183,865 & 467,570 \\
MAE (small) & 4.00e-9 & 9.46e-12 & 3.31e-15 & 1.09e-16 \\
RMSE (small) & 6.69e-8 & 8.83e-11 & 2.36e-11 & 1.05e-14 \\
SSPI & 0.8952 & 0.6900 & 0.9964 & 0.9874 \\
Max abs error & 2.82e-5 & 3.32e-8 & 1.72e-7 & 5.14e-12 \\
\midrule
Num. other pred. & 410,816,135 & 2,369,558 & 410,816,135 & 2,369,558 \\
MAE (other) & 6.51e-8 & 6.83e-8 & 1.56e-7 & 1.37e-7 \\
RMSE (other) & 3.71e-7 & 5.40e-7 & 1.01e-6 & 8.84e-7 \\
$R^2$ (other) & 0.999977 & 0.999530 & 0.999827 & 0.998741 \\
a1-index & 0.5211 & 0.2604 & 0.3873 & 0.2353 \\
a5-index & 0.7579 & 0.4732 & 0.8358 & 0.5190 \\
a10-index & 0.8164 & 0.5515 & 0.9193 & 0.6285 \\
a20-index & 0.8582 & 0.6245 & 0.9604 & 0.7178 \\
a100-index & 0.9179 & 0.7483 & 0.9939 & 0.9599 \\
\bottomrule
\end{tabular}
\end{table}

\begin{figure}[h]
    \centering
    \includegraphics[width=0.98\textwidth]{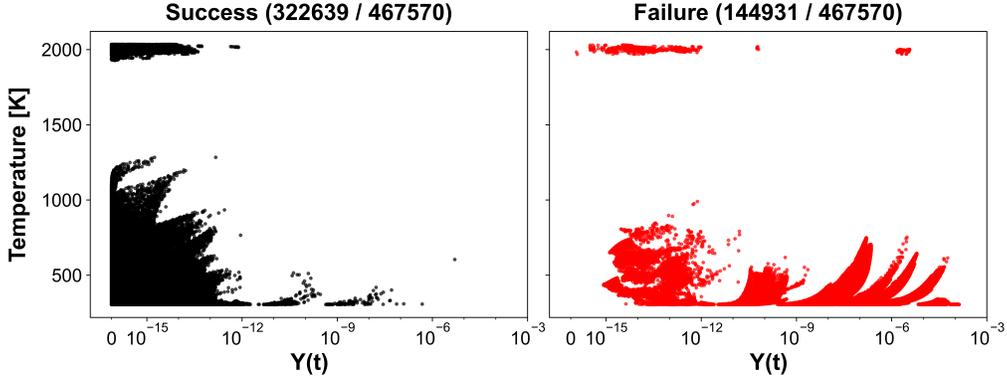}
    \caption{
    Scatter plots of BC model predictions for targets with magnitudes less than \( 10^{-15} \). The initial species mass fraction \( Y(t) \) prior to the reaction step is plotted against the initial temperature. Data points are grouped based on whether the model prediction successfully (left panel) or failed (right panel) to fall below \( 10^{-15} \). The counts of data points in each group are indicated above each panel. For both x-axes, values below \( 10^{-15} \) are represented on a linear scale, while values exceeding \( 10^{-15} \) are depicted on a logarithmic scale.
    }
    \label{fig:small_scale_error}
\end{figure}

Table \ref{tab:model_performance} summarizes the performance metrics for both models on the training and test datasets. Notably, the PT model maintains a remarkably high SSPI for both datasets, indicating a robust ability to predict extremely small concentrations. On the test set, the maximum absolute error for the smallest-scale targets is approximately $5.14 \times 10^{-12}$, which is acceptable given the context of species concentration magnitudes. Conversely, the BC model exhibits a decline in SSPI when transitioning from training to testing, with larger maximum errors for small targets.

This performance disparity can be attributed to the nature of the PT formulation. By applying a nonlinear power transformation, the PT model effectively "decompresses" the low-magnitude regime, simplifying the prediction task in this challenging region. Consequently, the PT model exhibits enhanced capacity for accurately and stably predicting tiny concentrations.

To further elucidate the limitations of the BC formulation, Figure \ref{fig:small_scale_error} presents scatter plots of the BC model’s predictions for targets with magnitudes below 10$^{-15}$. The initial species mass fraction prior to the reaction step 
\( Y(t) \) is plotted against the initial temperature, with data points grouped according to whether the prediction successfully (left panel) or failed (right panel) to fall below 10$^{-15}$. Failures predominantly involve states where 
\( Y(t) \) exceeds 10$^{-15}$, whereas successful predictions are mostly associated with states where 
\( Y(t) < 10^{-12} \). 
This discrepancy arises from the non-monotonic and nonlinear dependence of the BC formulation on the target concentration change \( \Delta Y \). As shown in Figure \ref{fig:pairity_plot_transform} a), the transformation’s behavior varies significantly with varying magnitudes of the species concentration before the reaction step. For targets with very small 
\( Y(t) \), the nonlinear mapping by the BC formulation effectively compresses the low-magnitude regime, making tiny changes more distinguishable and thus easier for the model to learn. This is evidenced by the red colored data points, representing burnt states with 
\( Y_{\text{NH}_3}(t) \) close to 0, whose mapped target value are distributed away from 0 in Figure \ref{fig:pairity_plot_transform} a) and yield prediction errors below 10$^{-8}$ (see c panel of Figure \ref{fig:pairity_plot_transform}). However, for states with larger 
\( Y(t) \), the transformation tends to map the targets close to zero, leading to a loss of distinguishability among different small changes. This is evidenced by the blue colored data points, representing cold mixture or unburnt states with larger 
\( Y_{\text{NH}_3}(t) \), whose mapped target value are pushed closer to zero, resulting in loss of distinguishability in Figure \ref{fig:pairity_plot_transform} a). This compression causes the model to struggle in learning accurate predictions for these states, and their prediction error are significantly larger, with errors reaching up to 10$^{-5}$ for targets around 10$^{-9}$ as in Figure \ref{fig:pairity_plot_transform} c).

In conclusion, the BC formulation is inherently limited in accurately predicting small-scale targets when the initial species mass fraction is not near zero. Its nonlinear, non-monotonic dependence on  
\( Y(t) \) causes effective decompression only in states with very small initial concentrations. For states with larger initial species mass fractions, the mapping fails to preserve the necessary distinctions, leading to unreliable predictions. Therefore, use of the BC-based model should be restricted to scenarios where the species concentrations are initially very low. As evidenced by the training dataset of the current study, these states are most likely to be cold mixture or unburnt states, where larger mass fractions of fuel or oxidizer species exist and the reaction process is slow. This corresponds to a common solution applied in previous works on deep learning for combustion chemistry integration, where either a frozen temperature limit ranging from 500 K to 900 K would be set and 
\( \Delta Y \) for states below the limit are set as 0 or obtained through traditional numerical integration instead. Although such strategies do not significantly impact statistical results in previous CFD studies, their applicability may be limited in cases where low-temperature chemistry critically influences the phenomena of interest. This work suggests that the formulation of the target  
\( \Delta Y \) substantially affects prediction reliability, and simply modifying the target formulation can mitigate these issues.

Performance metrics for prediction targets exceeding the small scale threshold are also shown in Table \ref{tab:model_performance}. The a20-index, a metric commonly used in machine learning, measures the percentage of predictions for which the absolute relative error does not exceed 20\%. Higher values indicate superior performance. Similarly, other a-indexes are defined to provide a broad overview of the relative error distribution. Although the PT model exhibits superior overall a-indexes for absolute relative errors greater than 5\%, it demonstrates a smaller a1-index compared to the BC model across both training and testing datasets. This indicates that the BC model achieves a higher proportion of predictions with an absolute relative error below 1\%. Given the extensive literature documenting successful applications of the BC-type target formulation, this approach has demonstrated its effectiveness. The BC model also exhibits notably lower MAE and RMSE values on both training and test datasets. Based on these observations, we conclude that relying exclusively on models trained with PT-formulated targets may not be optimal for subsequent CFD simulations. Instead, we propose a hybrid approach that employs BC models for thermochemical states with temperatures above 1000 K and PT models for states below this threshold. The choice of 1000 K is supported by Figure \ref{fig:small_scale_error}, which indicates that most prediction failures of the BC model for small-scale targets occur at temperatures below this value.

Furthermore, it is shown that the R$^2$ metric may be less informative when dealing with multiscale targets in combustion chemistry integration. Excluding small-scale targets from the calculation of R$^2$ has little effect on its value, and all models achieve very high R$^2$ scores on both training and testing datasets.

\subsection{Application to Premixed Flame in Two-Dimensional Homogeneous Isotropic Turbulence}
The previously trained DFODE models were applied to 2D HIT flame simulations. Results from two inference runs using the DFODE models were compared against reference simulations utilizing CVODE for detailed chemistry integration. One simulation employed solely the baseline BC model, while the other combined the BC model for states above 1000 K with the PT model for states between 305 K and 1000 K. For temperatures below 305 K, the changes in species concentrations 
\( \Delta Y \) were set to zero. All simulations were conducted over a duration of 4 ms, during which the flame kernel propagated near the boundary.

\begin{figure}[h]
    \centering
    \includegraphics[width=0.75\textwidth]{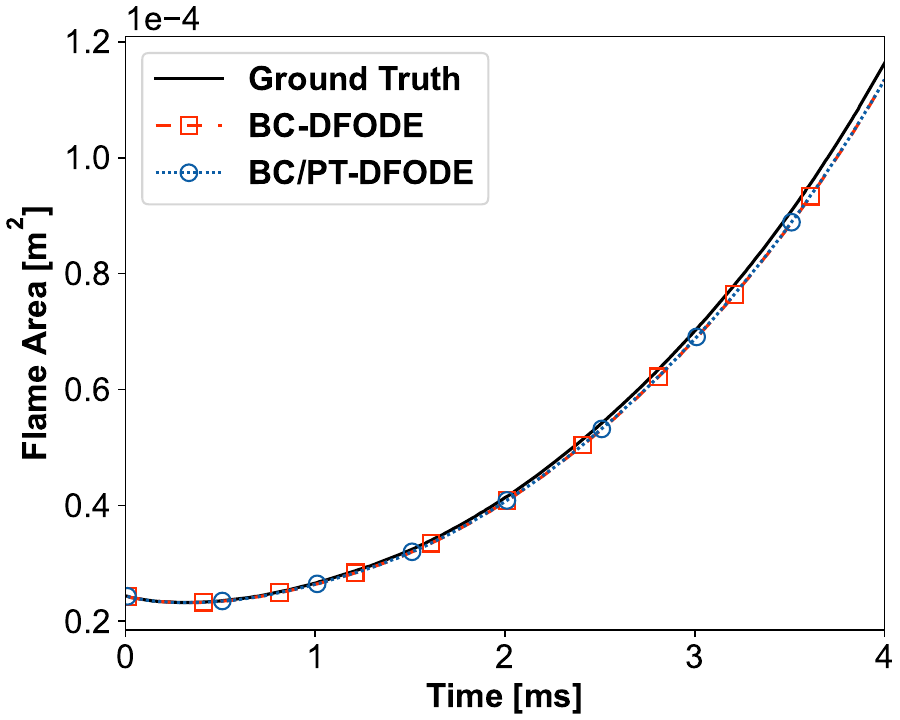}
    \caption{
    Temporal evolution of flame area in 2D HIT simulations: comparison between CVODE reference solution and DFODE inference runs.
    }
    \label{fig:2d_hit_flame_area}
\end{figure}

\begin{figure}[h]
    \centering
    \includegraphics[width=0.98\textwidth]{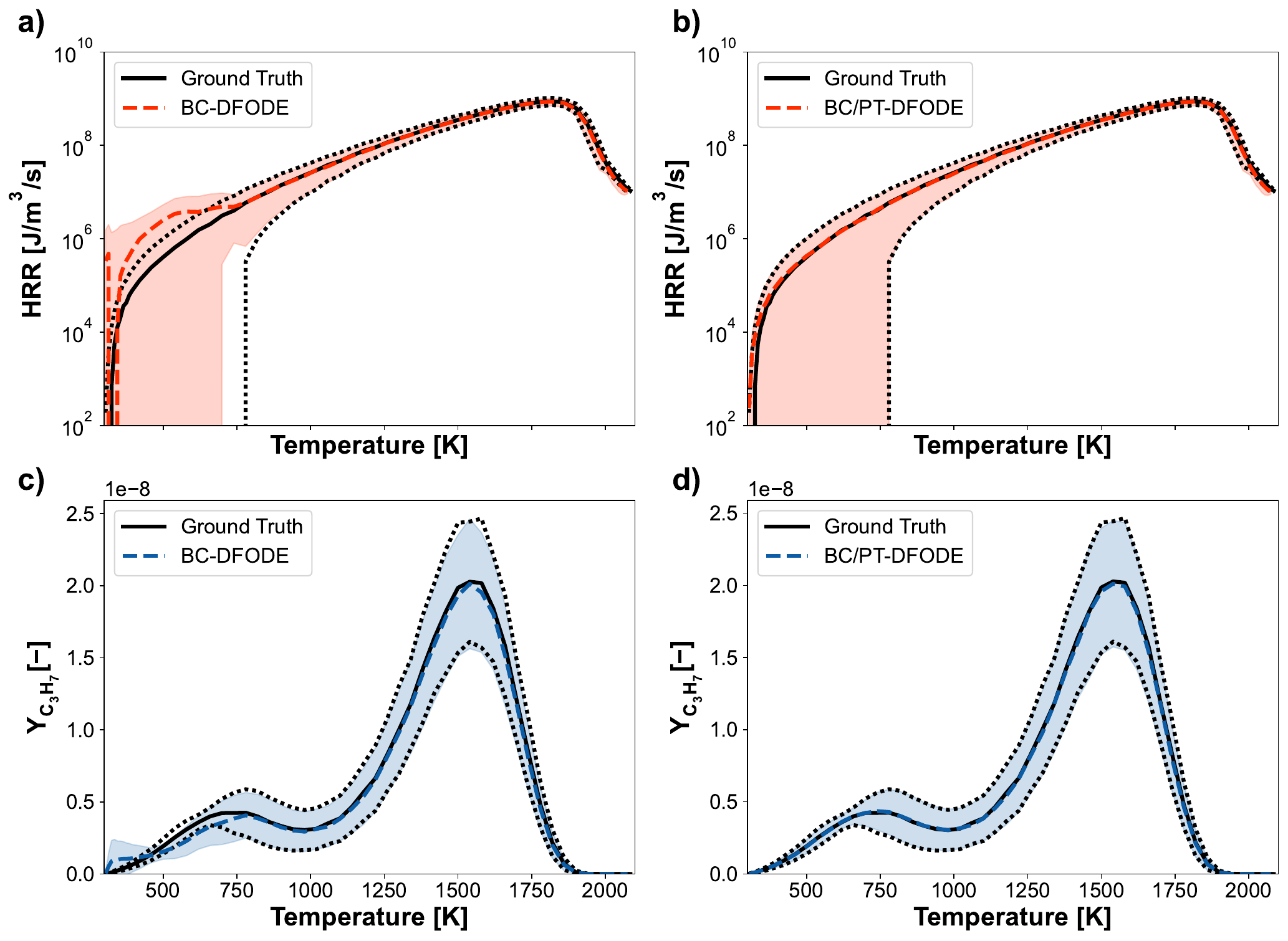}
    \caption{
    Conditional-averaged heat release rate (HRR) and C$_3$H$_7$ mass fractions as functions of temperature in 2D HIT simulations at the 4 ms snapshot. Panels a) and c) show results for the BC-DFODE model simulation, while panels b) and d) display results for the BC/PT-DFODE model. The shaded regions represent the mean $\pm$ one standard deviation for the DFODE runs, whereas the mean $\pm$ one standard deviation for the CVODE results is indicated by dotted lines.
    }
    \label{fig:2d_hit_condi_avg}
\end{figure}

\begin{figure}[p]
    \centering
    \includegraphics[width=0.98\textwidth]{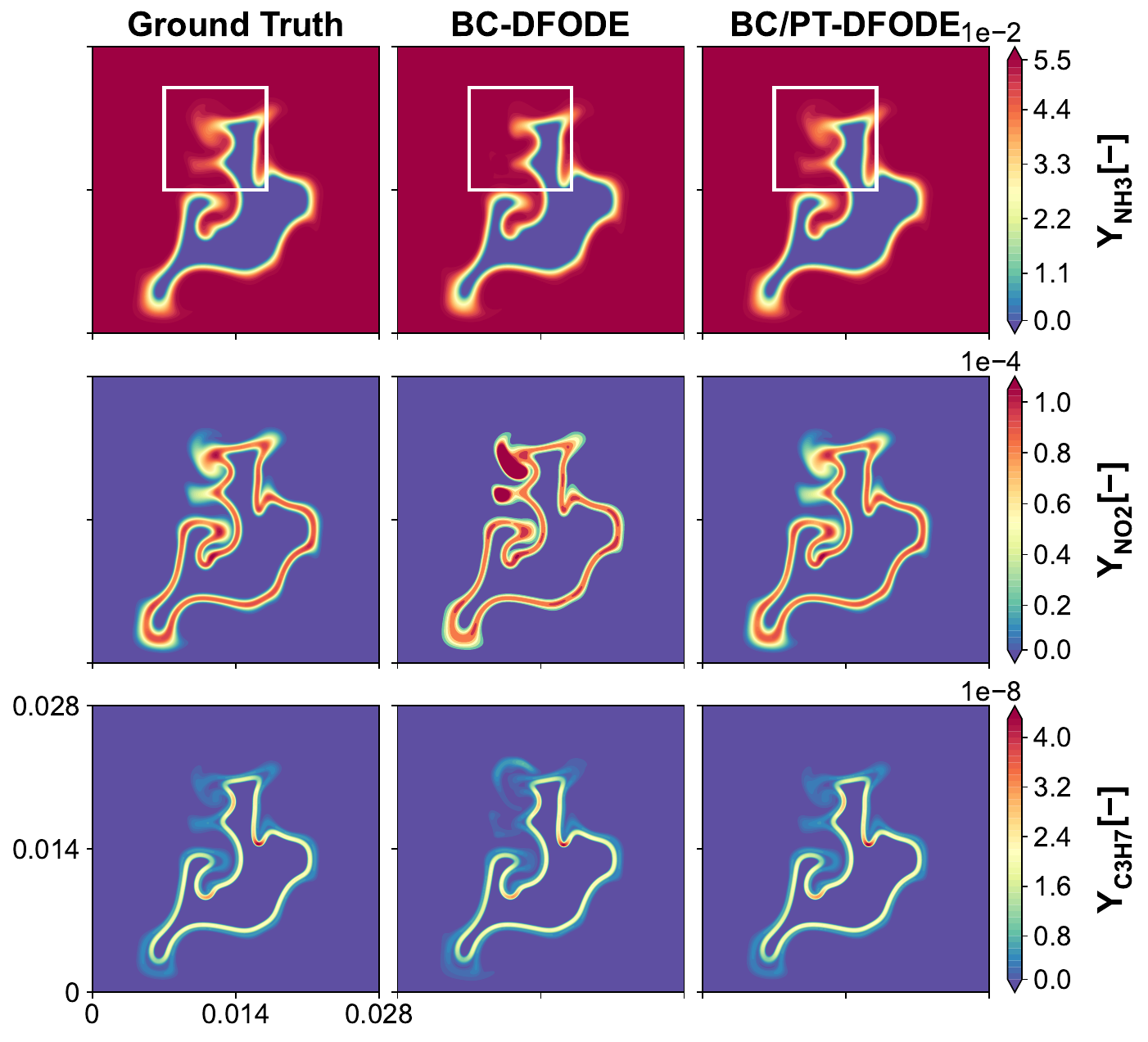}
    \caption{
    Contour plots of species mass fractions (NH$_3$, NO$_2$, and C$_3$H$_7$) in 2D HIT simulations at the 4 ms snapshot, comparing results from CVODE (left column) and DFODE simulations (middle and right columns). The top row shows NH$_3$, the middle row NO$_2$, and the bottom row C$_3$H$_7$, with colorbars indicating the respective species concentrations.
    }
    \label{fig:2d_hit_contour_comp}
\end{figure}

Figure \ref{fig:2d_hit_flame_area} illustrates the temporal evolution of the flame area in the 2D HIT simulations, providing an overview of flame kernel development. Both DFODE inference runs, one using only the BC model and the other combining BC and PT models, show excellent agreement with the CVODE reference solution, with their trajectories nearly indistinguishable throughout the entire simulation duration. This indicates that the DFODE models used for combustion chemistry integration are sufficiently accurate in capturing the overall dynamics of the reacting flow. Additionally, the close correspondence between the two DFODE-based results suggests that the optimizations designed for small-scale predictions in the low-temperature regime do not significantly compromise the models’ ability to accurately represent large-scale statistical features of the combustion process.

Figures \ref{fig:2d_hit_condi_avg} and \ref{fig:2d_hit_contour_comp} illustrate the impact of employing PT-formulated DFODE models for chemistry integration in the low-temperature regime. Specifically, Figure \ref{fig:2d_hit_condi_avg} shows the conditional-averaged heat release rate (HRR) and C\(_3\)H\(_7\) mass fractions as functions of temperature. Results indicate that relying solely on the BC model leads to significant deviations in both HRR and C\(_3\)H\(_7\) concentrations below 1000 K. Conversely, integrating the PT-based DFODE models for states under 1000 K yields results that closely align with the CVODE reference, demonstrating improved accuracy in capturing low-temperature chemical kinetics during pre-ignition and ignition phases.

Similarly, the spatial distributions of species at the 4 ms snapshot, shown in Figure \ref{fig:2d_hit_contour_comp}, reveal that the BC model inadequately reproduces the spatial patterns of species such as NH\(_3\), particularly in mixing regions on the unburnt side of the flame front. This deficiency is more pronounced for intermediate and smaller species, such as NO\(_2\), where the BC model fails to capture the distribution observed in the CVODE solutions. In contrast, the combined approach employing both BC and PT models achieves close agreement with CVODE results for NH\(_3\), NO\(_2\), and C\(_3\)H\(_7\), accurately reproducing even the fine-scale structures observed in the species distributions.

These findings demonstrate that the DFODE-based chemistry integration models, especially when incorporating the PT formulation for low-temperature regimes, can reliably replicate the complex interactions between turbulence and chemistry in 2D HIT flames. The models' ability to accurately predict both global flame features and local species distributions underscores their potential for efficient and precise simulations of turbulent combustion systems. This capability is particularly relevant for modeling ammonia fuel blends, where flame broadening and extinction phenomena are influenced significantly by low-temperature chemistry and the associated regimes.

The simulations were conducted using 32 Intel Xeon Gold 6330 CPU cores, with an additional 2 NVIDIA GeForce RTX 4090 GPUs used for DFODE inference. Compared to previous works \cite{li_comprehensive_2025}, data transfer between CPU and GPU has been optimized through data batching, which simplified the process of invoking PyTorch for DFODE inference and improved data transfer speed as well as memory consumption, allowing the solver to perform better when processing large-scale data. As a result, DFODE simulations achieved a 526 times speedup in chemistry source term calculations. When considering the overall calculation, this translates to a total speedup of 20 times compared to the simulation using CVODE.

\section{Conclusions}
This study successfully extended and validated a deep learning framework aimed at accelerating chemical kinetics in simulations of ammonia and natural gas combustion. The work addressed two key challenges in developing robust DFODE surrogate models: the quality of training data and the accurate prediction of multiscale target variables.

First, a physics-informed data augmentation strategy was introduced to enhance the training dataset generated from a one-dimensional laminar flame. This approach combines flame structure interpolation with constrained random perturbations, effectively alleviating data imbalance in highly reactive zones and expanding the coverage of the thermochemical state space. Validation in 1D flame simulations confirmed that models trained with this augmented data achieved significantly improved fidelity. They accurately reproduced temperature profiles, heat release rates, and the concentrations of both major and intermediate species, which was not attainable with the original data.

Second, this work evaluated transformation strategies for modeling multiscale species concentration changes. It was demonstrated that target formulations used in prior studies struggle to reliably predict extremely small concentration changes in low-temperature regimes. This limitation can compromise the physical consistency of the simulations. To address this, a power transformation directly applied to the concentration change was proposed. This scale-separation approach significantly improved prediction stability and accuracy for targets with low magnitudes. A hybrid modeling strategy was developed that leverages the high accuracy of the original target formulation at high temperatures and the stability of the power transformation at low temperatures. This combined approach provided a comprehensive and robust solution.

The effectiveness of the entire framework was validated through a posteriori testing in a two-dimensional turbulent flame simulation. The hybrid DFODE model accurately captured both the overall flame evolution and the detailed spatial distributions of species in low-temperature mixing regions. The results showed excellent agreement with the high-fidelity CVODE reference solution. Importantly, this level of accuracy was achieved with a substantial performance gain, delivering up to a twentyfold speedup in the overall simulation time.

In conclusion, this study presents a scalable and robust methodology for applying deep learning to complex chemical kinetics in numerical combustion modeling. The proposed data augmentation and scale-separation strategies provide a clear pathway for developing accurate and efficient surrogate models for high-fidelity combustion simulations. This work not only advances machine learning applications in ammonia combustion modeling but also offers a versatile framework that can be extended to other complex fuels, thereby facilitating the accelerated development and optimization of zero-carbon energy technologies.

\section*{Acknowledgements} \addvspace{10pt}

This work is supported by the National Natural Science Foundation of China (Grant Nos. 92270203, 52276096, and 52441603) and the China Postdoctoral Science Foundation (Grant No. 2023M730040).


\bibliographystyle{ama} 
\bibliography{library_nh3ch4_dnn}



\end{document}